\documentclass[aps,twocolumn,groupedaddress,showpacs]{revtex4}
\usepackage{graphicx,amssymb,amsbsy,color}
\input epsf 
\epsfverbosetrue 

\begin{document}

\title{Using modified Gaussian distribution to study the physical properties of
one and two-component ultracold atoms }
\author{Chou-Chun Huang and Wen-Chin Wu}
\affiliation{Department of Physics, National Taiwan Normal
Univesity, Taipei 11650, Taiwan}

\date{\today}

\begin{abstract}
Gaussian distribution is commonly used as a good approximation to
study the trapped one-component Bose-condensed atoms with
relatively small nonlinear effect. It is not adequate in dealing
with the one-component system of large nonlinear effect, nor the
two-component system where phase separation exists. We propose a
modified Gaussian distribution which is more effective when
dealing with the one-component system with relatively large
nonlinear terms as well as the two-component system. The modified
Gaussian is also used to study the breathing modes of the
two-component system, which shows a drastic change in the mode
dispersion at the occurrence of the phase separation. The results
obtained are in agreement with other numerical results.
\end{abstract}

\maketitle

\section{Introduction}
\label{sec1}

Single-component Bose-Einstein condensate (BEC) in trapped atoms
has been observed more than ten years ago. Since then, in
particular in recent years, laser cooling technique has been
extended to trap atoms of multiple components. The basic technique
is called ``sympathetic cooling" that cools the second kind of
atoms by means of the first kind pre-cooled atoms. Various
combinations of two-component atomic gas have been mixed and
cooled. The first was done on the mixture of the same kind of
bosonic atoms with different hyperfine spin states. For instance,
the combinations of $^{87}$Rb$|2,2\rangle$ and
$^{87}$Rb$|1,-1\rangle$ states \cite{myatt97}, and of
$^{87}$Rb$|2,1\rangle$ and $^{87}$Rb$|1,-1\rangle$ states
\cite{hall98}. Soon after, the experiment has been extended to
cool the mixture of different kinds of bosons. This includes the
mixture of $^{87}$Rb and $^{41}$K atoms \cite{modugno01} and
$^{133}$Cs and $^{87}$Rb atoms \cite{anderlini05}. The mixtures of
bosons and fermions have also been achieved, such as the
combination of $^{7}$Li and $^{6}$Li atoms
\cite{truscott01,schreck01}, $^{6}$Li and $^{23}$Na atoms
\cite{hadzibabic02}, and $^{40}$K and $^{87}$Rb atoms
\cite{goldwin02}. The fermion-fermion mixtures have also been
accomplished, such as the combinations of two different spin
states in $^{40}$K atoms \cite{regal03} and in $^{6}$Li atoms
\cite{jochim03,strecker03}. The latter has led to the formation of
bosonic molecules using Feshbach resonance and has advanced the
observation of BEC-BCS crossover
\cite{chin04,partridge05,zwierlein05}. More recently, imbalanced
spin population in fermion-fermion mixture has also been created
and boosted researches on these systems
\cite{zwierlein06,partridge06}.

In the literature, there exists theoretical works on equilibrium
and dynamical properties of the two-component cold atoms,
including the investigations on the boson-boson mixtures
\cite{ho96,bashkin97,pu98,molmer98,nygaard99} and the
fermion-fermion mixtures \cite{kinnunen06,pieri06}. For
two-component boson-boson mixtures, in which both components are
Bose condensed, there are two major theoretical approaches. One is
concerned with the Thomas-Fermi approximation (TFA), in which
kinetic-energy terms are ignored. The other starts with the
coupled Gross-Pitaveskii (GP) equations, which are then solved
numerically.
In this paper, we attempt an alternative approach to investigate
the two-component system based on variational method. Through the
introduction of a modified Gaussain trial function, one is able to
study the condition of the system being in the state of phase
separation or miscible. The breathing modes of the system is also
studied and shown to be intimately connected to their ground-state
behavior. As a matter of fact, the modified Gaussian distribution
is clearly shown to be a better description than the simple
Gaussian even to the one-component trapped BEC, especially when
the nonlinear effect is more important. Variational method in
adjunction with the introduced trial modified Gaussian function
can be easily generalized to study the two-component boson-fermion
and fermion-fermion mixtures. The latter includes the system of
imbalanced spin population of current interest.

The paper is organized as follows. In Sec.~\ref{sec2}, we first
give the Hamiltonian of a two-component boson-boson mixture.
Coupled Gross-Pitaveskii energy functionals are derived for the
system assuming that both components are fully Bose condensed. A
subsection is devoted to sketch the variational method. In
Sec.~\ref{sec3}, we introduce a trial ``modified Gaussian
function" whose ground-state behaviors to the one-component Bose
system are compared in details with those of the simple Gaussian
distribution and the TFA. In Sec.~\ref{sec4}, the trial modified
Gaussian function are applied to study the equilibrium and
dynamical properties of the two-component Bose system, and in
Sec.~\ref{sec5}, a brief conclusion is given.
In the current context, we focus on the trapped system of
boson-boson mixture. Similar approaches to the fermion-boson and
fermion-fermion mixtures will be possible future works.

\section{Formalism and Methodology} \label{sec2}

\subsection{Coupled GP Energy Functionals}
\label{sec21}

For a trapped two-component boson-boson mixture, the Hamiltonian
can be given by
\begin{eqnarray} H=H_1+H_2+H_{12},\label{eq:H1}
\end{eqnarray}
where
\begin{eqnarray}
H_i&=&\int d{\bf r} \phi^\dagger_i({\bf
r})\left[-{\hbar^2\nabla^2\over
2m_i}+V_i({\bf r})-\mu_i\right]\phi_i({\bf r})\nonumber\\
&+& {g_{ii}\over 2}\int\int d{\bf r} d{\bf r}\prime
\phi^\dagger_i({\bf r}) \phi^\dagger_i({\bf r}^\prime)\delta({\bf
r}-{\bf r}^\prime)\phi_i({\bf r}^\prime)\phi_i({\bf r})
\label{eq:H2}
\end{eqnarray}
and
\begin{eqnarray} H_{12}&=&g_{12}\int\int d{\bf r} d{\bf r}^\prime
\phi^\dagger_1({\bf r})
\phi^\dagger_2({\bf r}^\prime)\delta({\bf r}-{\bf
r}^\prime)\phi_2({\bf r}^\prime)\phi_1({\bf r}). \label{eq:H3}
\end{eqnarray}
Here $i=1,2$ denote the species $i$ of the mixture,
$\phi_i^\dagger({\bf r})$ and $\phi_i({\bf r})$ are creation and
annihilation operators on the particle of species $i$ at ${\bf
r}$, and $\mu_i$ denote the chemical potential of the species $i$.
The entire system is trapped in an isotropic magnetic potential
$V_i({\bf r})={1\over 2}m_i\omega_i^2 r^2$ respectively for each
component $i$. It is assumed that the interaction is dominated by
the $s$-wave approximation in the dilute limit. Consequently the
interaction coupling constants in (\ref{eq:H2}) and (\ref{eq:H3})
can be written in a unified way
\begin{eqnarray} g_{ij}={4\pi\hbar^2 a_{ij}\over m_{ij}}
, \label{eq:gs}
\end{eqnarray}
where $a_{ij}$ is the scattering length for a particle in species
$i$ with a particle in species $j$. The ``mass" $m_{ij}\equiv 2m_i
m_j /(m_i+m_j)$ with $m_i$ the bare mass of the particle in
species $i$. For simplicity, we shall denote $g_{11}\equiv g_1$;
$g_{22}\equiv g_2$ and $a_{11}\equiv a_1$; $a_{22}\equiv a_2$
afterwards.

At sufficiently low temperatures, consider that both components
are fully Bose condensed, the thermal average of $H_i$ in
(\ref{eq:H2}) then reduces to the Gross-Pitaveskii (GP) energy
functional:

\begin{eqnarray}
E_i=\langle H_i\rangle&=&\int d{\bf r} \Phi^*_i({\bf
r})\left[-{\hbar^2\nabla^2\over 2m_i}+{1\over 2}m_i\omega_i^2
r^2-\mu_i \right.\nonumber\\
&+&\left.{g_i\over 2}|\Phi_i({\bf r})|^2\right]\Phi_i({\bf r}),
\label{eq:E1}
\end{eqnarray}
where $\Phi_i=\langle \phi_i\rangle$. Similarly the thermal
average of $H_{12}$ in (\ref{eq:H3}) reduces to

\begin{eqnarray}
V_{12}=\langle H_{12}\rangle=g_{12}\int d{\bf r} |\Phi_1({\bf
r})|^2 |\Phi_2({\bf r})|^2. \label{eq:E2}
\end{eqnarray}

\subsection{Variational Method}
\label{sec22}

One major goal of this paper is to apply the variational method to
investigate the dynamics of the coupled two-component ultracold
atoms. The key to the variational approach is to first look for a
trial wave function $\Phi_i({\bf r})$ for each component $i$,
which is reasonable to the ground state of the system. With these
trial wave functions, one then incorporates appropriate dynamical
variables [$\Phi_i({\bf r})\rightarrow \Phi_i({\bf r},t)$], which
are sensible to the dynamics under studies. The next step is to
derive the corresponding Lagrangian: $L=\int d{\bf r} T-E$ using
$\Phi_i({\bf r},t)$. Here

\begin{eqnarray}
T=\sum_{i=1,2}{i\hbar\over 2}\left(\Phi_i^*{\partial \Phi_i\over
\partial t}-\Phi_i{\partial \Phi_i^*\over \partial t}\right)
\label{eq:T1}
\end{eqnarray}
and
\begin{eqnarray}
E=\langle H\rangle=E_1+E_2+V_{12}, \label{eq:T2}
\end{eqnarray}
where $E_1$, $E_2$, and $V_{12}$ are calculated based on
(\ref{eq:E1}) and (\ref{eq:E2}) with $\Phi_i({\bf r})$ being
replaced by $\Phi_i({\bf r},t)$ now. Finally a set of coupled
dynamical equations of motion can be derived through the
Euler-Lagrange equations:
\begin{eqnarray}
{\partial L\over \partial \beta}={d\over dt}{\partial L\over
\partial \dot{\beta}}, \label{eq:T3}
\end{eqnarray}
where $\beta$ is any one of the dynamical variables chosen. The
corresponding dynamics, in particular the collective mode
dispersion relation, can be obtained by solving the roots of the
dynamical matrix (diagonalization).

\section{Trial wave function}
\label{sec3}

Before a proper trial wave function is introduced for each species
in the trapped two-component cold atoms, it is useful to first
examine the validity of the simple Gaussian distribution.

\subsection{Remarks: Gaussian Distribution} \label{sec31}

When temperature approaches zero, single-component trapped Bose
condensate (isotropic case) is governed by the GP energy
functional given in Eq.~(\ref{eq:E1}). [In this case the index $i$
in (\ref{eq:E1}) is redundant and is thus removed.] When the
coupling $g$ is infinitesimally small, the ground state of the
system is perfectly a Gaussian distribution:
\begin{eqnarray}
\Phi({\bf r})&=& C\exp \left(-Ar^2 \right). \label{eq:gaussian}
\end{eqnarray}
Gaussian function is ``non-perturbative" and has the advantage of
easily integrated over the space. When $g$ starts to increase,
however, a Gaussian may no longer be a good one to describe the
system. In the limit of large $g$, Gaussian distribution fails and
the most common approximation to the ground state of the system is
the so-called ``Thomas-Fermi approximation" (TFA), which ignores
the kinetic term in (\ref{eq:E1}). The validity of the Gaussian
distribution upon the increase of $g$ can be sorted out as
follows. First, let's rescale the GP energy functional:
 $E/(\hbar\omega ) \rightarrow E$, $t\omega \to t$,
$r/\ell  \to r$ with $ \ell  \equiv \sqrt {{\hbar }/{m\omega }}$
being the length scale throughout this paper. Next, divide the
energy functional by $N$ (total number of particles),
Eq.~(\ref{eq:E1}) for the current one-component system then
becomes
\begin{eqnarray}
E/N=\int d{\bf r}{\Phi^* \left[ {-\frac{1}{2}\nabla ^2  +
  \frac{1}{2}r^2+\frac{m}{{2\ell}}Ng|{\Phi}|^2} \right] \Phi}.
\label{eq:E3}
\end{eqnarray}
By writing this way, it is seen clearly that in addition to $g$,
both the magnetic trap potential (through $\ell$) and $N$ are also
coupled to the nonlinear term. As a consequence, whether a
Gaussian is a good approximation will depend on the magnitude of
the front factor for the nonlinear term, $mNg/2\ell\propto
Ngm^{3/2}\sqrt{\omega}$. Provided that $g$ is fixed, the smaller
the value of $N\sqrt{\omega}$, the better the Gaussian
distribution is for the same atoms.

In current experiments performed on the single-component system,
nonlinear effect is typically not small. Application of the
Gaussian distribution to a real system thus always poses a
question mark. Moreover, in a system of two-component mixture, the
phenomenon ``phase separation" could exist when the interspecies
coupling $g_{12}$ is strong. In this case, using a Gaussian to
describe the distribution of the individual component will be
fundamentally ineffective.

\subsection{Modified Gaussian Distribution}
\label{sec32}

\begin{figure}
\vspace{-0.5cm}
\includegraphics[width=0.53\textwidth]{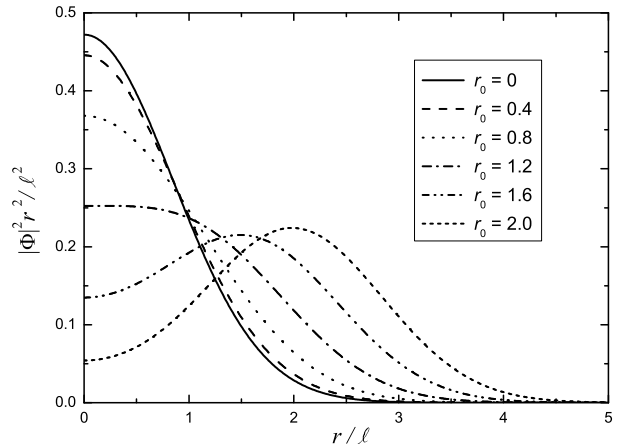}
\vspace{-1.0cm} \caption{The normalized density profile,
$|\Phi({\bf r})|^2r^2$, with $\Phi$ given by the modified Gaussian
function in (\ref{eq:phi1}). Here $A=0.8$ is fixed for all cases.}
\label{fig1}
\end{figure}

To better describe the system and for capturing the advantage of
easy integration of the Gaussian function, we propose the
following ``modified Gaussian distribution":
\begin{eqnarray}
\Phi({\bf r})&=& C\left\{\exp \left[-{A(r-r_0)^2\over 2} \right] +
 \exp \left[-{A(r+r_0)^2\over 2} \right] \right\}\nonumber\\
&=& C^\prime\exp \left(-{Ar^2\over 2} \right)\cosh(A r_0 r),
\label{eq:phi1}
\end{eqnarray}
which is the superposition of two Gaussian functions, with centers
deviated from the origin by $r_0$ and $-r_0$ respectively. As
written explicitly in the second line of (\ref{eq:phi1}), the
resulting function is a Gaussian function {\em modified}
(multiplied) by a hypercosine function.
In (\ref{eq:phi1}), $C$ and $C^\prime=2C\exp(-Ar_0^2/2)$ are
constants for normalization. $A$ and $r_0$ are parameters
corresponding to the amplitude and overall density profile of the
atom cloud. For a given $C$, the values of $A$ and $r_0$ are
determined upon minimizing the ground-state energy $E$ of the
system. In Fig.~\ref{fig1}, $r_0$ dependence of the normalized
density profile $|\Phi({\bf r})|^2r^2$ is shown with a fixed $A$.
As can be seen clearly, the maximum of $|\Phi({\bf r})|^2r^2$ has
a tendency to move outwards when $r_0$ is increased. When $r_0$ is
large, the maximum could move significantly away from the origin.
Having this feature, which is important when phase separation
occurs, the modified Gaussian function is considered to be more
effective for a two-component mixture.

\begin{figure}
\vspace{-1.0cm}
\includegraphics[width=0.53\textwidth]{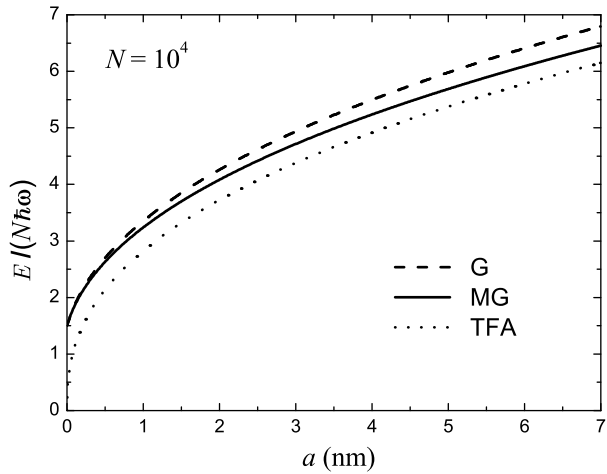}
\vspace{-1.0cm} \caption {Ground-state energy $E$ divided by
$\hbar\omega$ ($\omega =$ trap frequency) versus the scattering
length $a$ for single-component condensed bosons. The curves are
for Gaussian (G), modified Gaussian (MG), and TFA. The parameters
are $\omega=2\pi\times 160$Hz and $N=10^4$.} \label{fig2}
\end{figure}

Compared to the simple Gaussian distribution
[Eq.~(\ref{eq:gaussian})], the modified Gaussian consists one more
parameter ($r_0$), that allows for a better description of the
ground-state equilibrium properties of the system. As an
illustration, in Fig.~\ref{fig2} we plot the ground-state energy
$E$ [Eq.~(\ref{eq:E3})] as a function of the scattering length $a$
for the modified Gaussian distribution. The parameters are taken
to be: $\omega=2\pi\times 160$Hz and $N=10^4$. Similar results for
the simple Gaussian distribution and the TFA are also shown for
comparison. In the small-$a$ regime (small nonlinear effect), the
curve of modified Gaussian merges to the one of simple Gaussian,
as it must be. Gaussian distribution is exact in the $a\rightarrow
0$ limit. The ground-state energy $E$ is seen to be much
underestimated for TFA at small $a$. In the intermediate $a$
($=2\sim5$) regime, nonlinear effect becomes more and more
important, the curve of modified Gaussian starts to deviate
(becomes lower) from the one of a simple Gaussian. This indicates
that the modified Gaussian distribution is more favored by the
system so long as the ground-state energy is concerned. It is
worth noting that most current experiments lie in this
intermediate $a$ regime. In the regime of large $a$, the results
of TFA should be the most accurate ones.

\begin{figure}
\vspace{-1.0cm}
\includegraphics[height=12cm,width=0.5\textwidth]{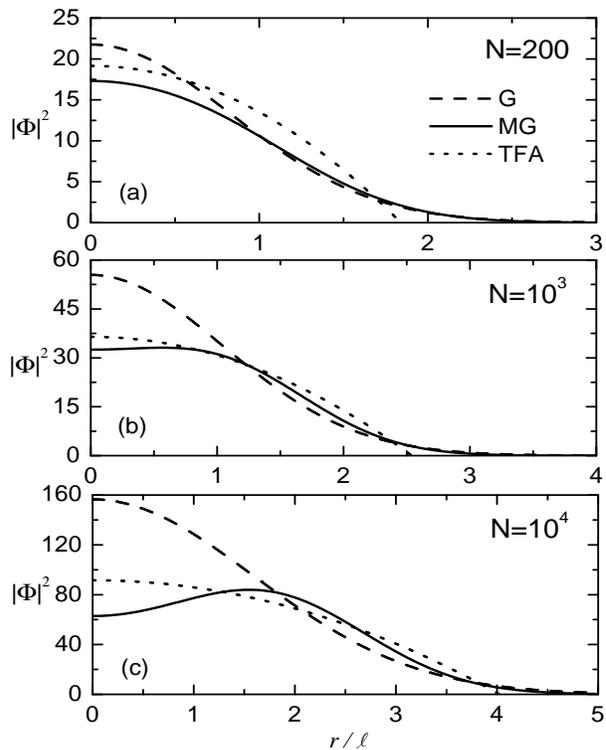}
\vspace{-1.5cm} \caption {The spatial dependence of the density
profile $|\Phi({\bf r})|^2$ for a one-component system. Three
kinds of distribution (Gaussian, modified Gaussian, and TFA) are
compared. Frame (a)-(c) are for number of atoms $N=200$, $10^3$,
and $10^4$ respectively.} \label{fig3}
\end{figure}

In Fig.~\ref{fig3}, spatial distributions (density profile) of the
three cases (Gaussian, modified Gaussian, and TFA) are compared
for different total numbers of particle $N$. The case of larger
$N$ corresponds to the case of larger nonlinear effect. In frame
(a) of smaller $N$, the curve of modified Gaussian is closer to
the one of simple Gaussian -- with the maximum at the origin. When
$N$ is larger [frame (b) \& (c)], the curves of modified Gaussian
behave closer to the ones of TFA, although the maxima may deviate
from the origin. As far as a real distribution is concerned, this
feature is somewhat defective. It should be emphasized, however,
that modified Gaussian has an overall distribution which is closer
to the one of TFA and is more accurate for large nonlinear effect.
Moreover, modified Gaussian inherits the advantage of easy
integration of simple Gaussian, which makes variational method
feasible for the problems.

In next section, equilibrium property and dynamics of a
two-component system are studied. Breathing modes of such system
will be shown to depend crucially on whether the system is
miscible or phase-separated.

\section{Two-component mixture}
\label{sec4}

The analyses in previous section are limited to single-component
systems. In the case of two-component mixture, the system could
undergo a phase transition from a miscible state to a
phase-separated state when the inter-species coupling is strong.
For current inhomogeneous but isotropic system, when phase
separation exists, one component of the atoms will have maximum
density deviated from the origin, while the density of the other
component remains maximum at the origin. As pointed out before, a
simple Gaussian is ineffective to describe such phenomenon. The
present section is devoted to study how the phase separation
occurs when the modified Gaussian distributions are in effect. At
the same time, the dynamics (breathing modes) of the two-component
system will also be studied starting from the same modified
Gaussian distribution.

\subsection{Equilibrium Property}
\label{sec41}

\begin{figure}
\vspace{-1.0cm}
\includegraphics[width=0.5\textwidth]{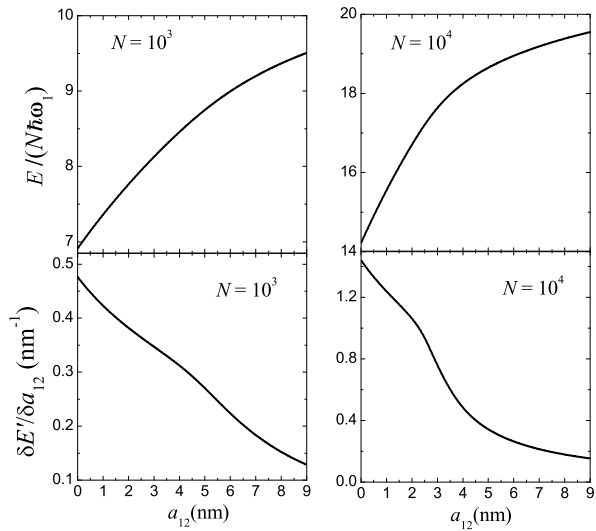}
\vspace{-1.0cm} \caption{The dependence of inter-component
scattering length $a_{12}$ on the ground-state energy $E$, divided
by $N\hbar \omega_1$, of a two-component Boson system. Modified
Gaussian distribution is undertaken. The slopes of $\partial
E/\partial a_{12}$ vs. $a_{12}$ are also shown [$E^\prime\equiv
E/(N\hbar\omega_1)$]. Left column corresponds to $N_1=N_2\equiv
N=10^3$, while right column corresponds to $N=10^4$. See text for
the values of other parameters.} \label{fig4}
\end{figure}

The ground-state properties of the two-component bosonic atoms are
first studied. Considering that both components are described by
the modified Gaussian distribution
\begin{eqnarray}
\Phi_i({\bf r})&=&C_i\exp\left({-\frac{1}{2}A_i r^2}
\right)\cosh(A_i r_i r), \label{eq:phi2}
\end{eqnarray}
where $i=1,2$ and the normalization factor $C_i$ is dependent of
the total number $N_i$ of individual component $i$. The
ground-state energy $E$ of the system can be calculated by
substitution of (\ref{eq:phi2}) into
Eqs.~(\ref{eq:H1})--(\ref{eq:E2}) and (\ref{eq:T2}). Given three
coupling strength $a_1$, $a_2$, and $a_{12}$, and the particle
number $N_1$ and $N_2$, final result of $E$ can be obtained by
minimizing $E$ against the variation of $A_1$ and $A_2$, and $r_1$
and $r_2$. In Fig.~\ref{fig4}, the results of $E$ are shown as a
function of the inter-component scattering length $a_{12}$. Two
cases of equal population $N_1=N_2\equiv N=10^3$ and $N=10^4$ are
presented. The slopes, $\partial E/\partial a_{12}$ vs. $a_{12}$
are also shown. For easy comparison, throughout this section the
parameters are taken to be the same as those in Ref.~\cite{pu98}.
Based on the experiment of Na-Rb mixture,  the scattering lengths
of Na and Rb atoms are chosen to be 3~nm and 6~nm respectively.
Regarding the external magnetic trap, the corresponding trap
frequencies are $\omega_1=2\pi\times 310Hz$ for Na and
$\omega_2=2\pi\times 160Hz$ for Rb respectively. In both plots in
Fig.~\ref{fig4} (especially $\partial E/\partial a_{12}$ vs.
$a_{12}$), one sees clear evidence that the system undergoes a
transition from one phase to another at $a_{12}\sim 2-2.5$ when
$N=10^4$ and at $a_{12}\sim 4-5$ when $N=10^3$.

\begin{figure}
\vspace{-0.5cm}
\includegraphics[height=5cm,width=0.5\textwidth]{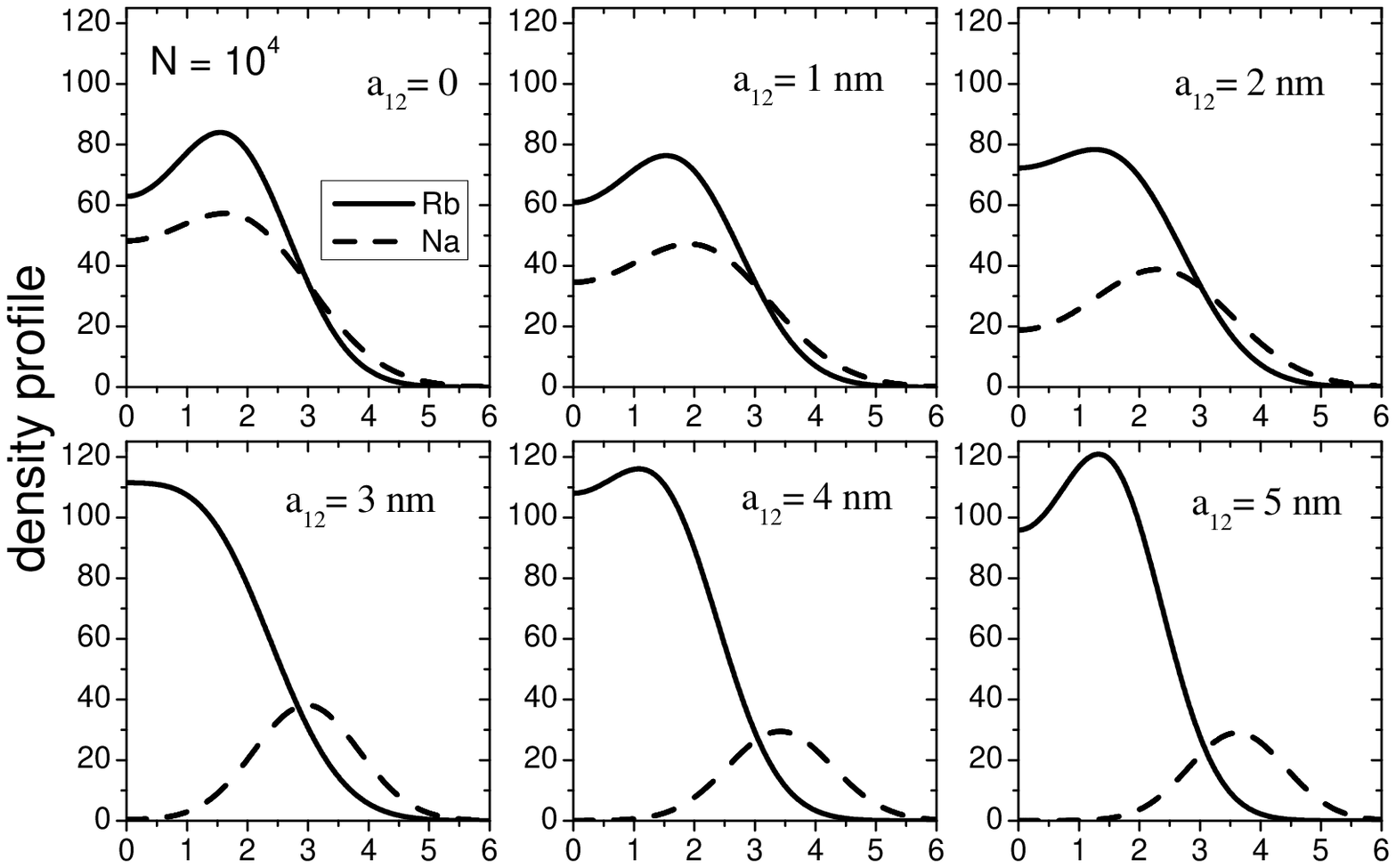}
\vspace{-0.5cm}
\includegraphics[height=5cm,width=0.5\textwidth]{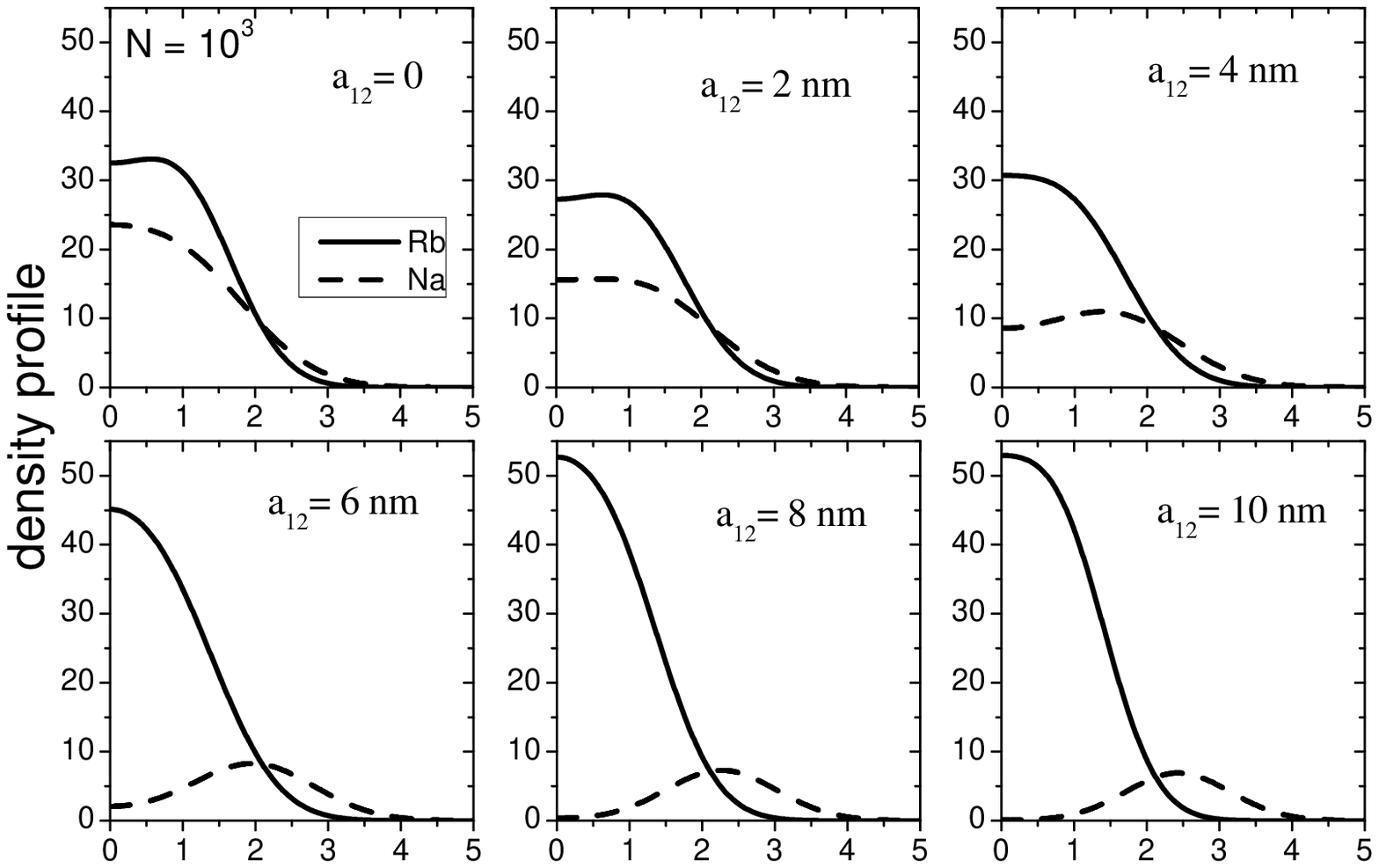}
\vspace{-0.5cm}
\includegraphics[height=5cm,width=0.5\textwidth]{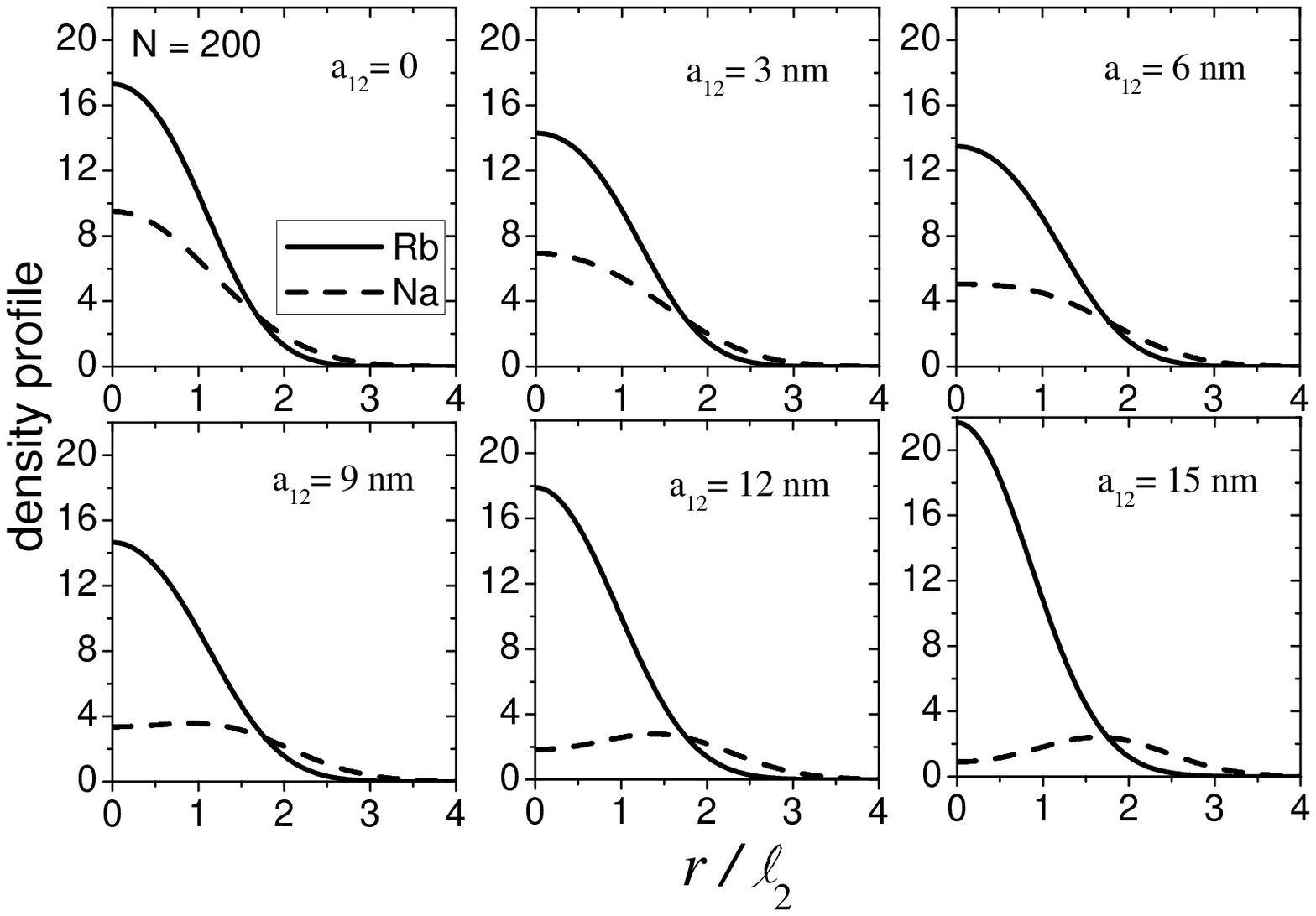}
\vspace{0.2cm}\caption {The individual density profile plotted in
a variety of the interspecies scattering length $a_{12}$ for a
two-component system. From the top to the bottom are for $N_1=
N_2\equiv N=10^4$, $10^3$ and $200$ respectively. Note that in $x$
axis, $r$ is scaled to $\ell_2=\sqrt {{\hbar }/{m_2\omega_2 }}$. }
\label{fig5}
\end{figure}

Fig.~\ref{fig5} shows how this transition behaves in the density
profiles of the two-component system. The figures are plotted in a
variety of interspecies scattering lengths ($a_{12}$) with fixed
$a_1=3$ nm and $a_2=6$ nm. In the case of $N=10^4$, it is found
that both $|\Phi_1(r)|^2$ and $|\Phi_2(r)|^2$ have maximum away
from the center when $a_{12}=0$. It has to be emphasized again,
that modified Gaussian distribution has a feature that the maximum
density could deviate from the origin due to large nonlinear
effect. Even though $a_{12}=0$, the individual nonlinear effect is
still large, which lead to such ``unphysical" result. As mentioned
before, it is the fact that modified Gaussian has an overall
distribution more close to TFA, that one has to look into. When
$a_{12}$ becomes larger ($a_{12}\sim 2-2.5$), the maximum of
$|\Phi_1(r)|^2$ is pushed significantly away from the center,
while the maximum of $|\Phi_2(r)|^2$ is right at the center -- the
pheomenon of phase separation.

\begin{figure}
\vspace{-0cm}
\includegraphics[width=0.55\textwidth]{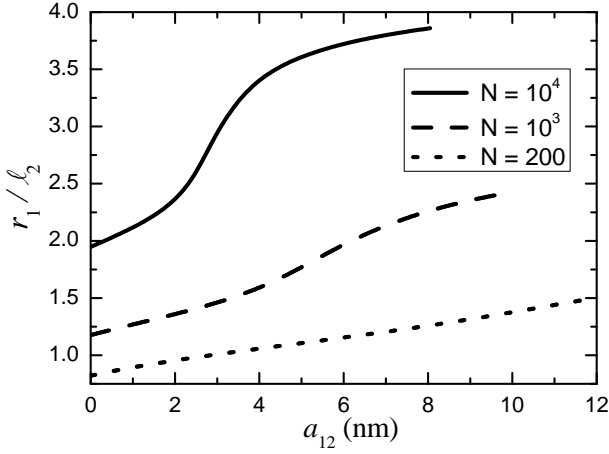}
\vspace{-0.5cm} \caption {The value of the parameter $r_1$ divided
by $\ell_2$ ($r_1$ is referred to Na component in Fig.~5) for the
modified Gaussian plotted as a function of the interspecies
scattering length $a_{12}$.} \label{fig6}
\end{figure}

To see this phenomenon more clearly, in Fig.~\ref{fig6} we plot
the value of parameter $r_1$ (for the modified Gaussian) versus
$a_{12}$. It is reminded that $r_1$ is referred to the Na
component in Fig.~5. In the case of $N=10^4$, there is a sudden
change for $r_1$ at $a_{12}\sim 2-2.5$. This is in great
consistence with the ground-state energy shown in Fig.~\ref{fig4}
where a sharp change also occurs at this range of $a_{12}$. In
view of the original definition of $r_1$ in the first line of
Eq.~(\ref{eq:phi1}) ({\rm i.e.}, $r_0$ there), a sudden change of
$r_1$ means a sudden change of the coordinate of density maximum
(see also Fig.~1). This sudden change indeed corresponds to the
transition of ``phase separation". In the case of $N=10^3$, we
also see a consistent picture among the results in
Figs.~\ref{fig4}--\ref{fig6}. A phase separation occurs at
$a_{12}\sim 4-5$. When $N=200$, the phenomenon is less obvious.
However, one can still see a phase separation that appears at
$a_{12}\sim 9-10$.

\subsection{Breathing Modes} \label{sec42}

In this section, we investigate how the phase separation affects
the dynamics of the two-component system. It was shown in
Ref.~\cite{pu98} that for a two-component mixture, the collective
mode dispersion will have a drastic change near the occurrence of
phase separation. Since the property of collective mode is
intimately connected to the equilibrium property of the system,
the sharp transition in the ground-state energy of the system
could imply a drastic collective mode dispersion change in the
same regime.

In the current context, we will focus on the breathing modes of
the system. One first extends the modified Gaussian distribution
for each component [Eq.~(\ref{eq:phi2})] to include the dynamical
variables \cite{martikainen04,huang05}:

\begin{eqnarray}
\Phi_i({\bf
r},t)&=&C_i\exp\left\{{-\frac{1}{2}A_i[1+\varepsilon_i(t)+
i\varepsilon^\prime_i(t)]r^2} \right\}\nonumber\\
&&\times \cosh(A_i r_i r). \label{eq:phi3}
\end{eqnarray}
Here $\varepsilon_i$ and $\varepsilon_i^\prime$ correspond to the
fluctuations of local amplitude and local phase of the atom cloud
along the radial $r$ direction. $A_i$ and $r_i$ are determined
earlier through minimizing the ground-state energy of the system
(see previous subsection). Since we are interested in the
breathing modes of the lowest energy and of only one radial node
($n=1$), $\varepsilon_i$ and $\varepsilon_i^\prime$ are coupled
only to the $r^2$ term in the Gaussain function in
(\ref{eq:phi3}). One may introduce extra dynamical variables to
the hypercosine function in (\ref{eq:phi3}), that could lead to
more branches of radial modes. Some of them will have more than
one node ($n \geq 2$).

\begin{figure}
\vspace{-0.5cm}
\includegraphics[width=0.5\textwidth]{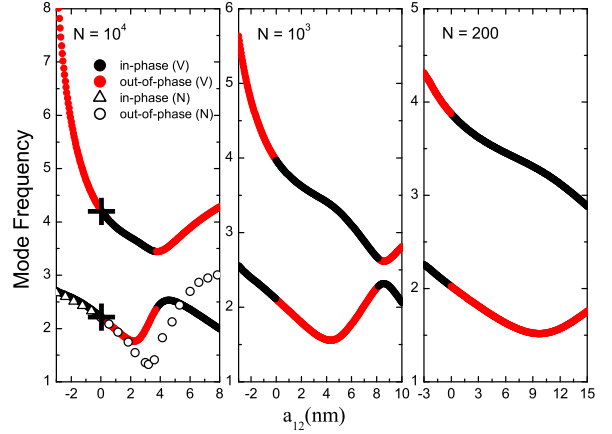}
\vspace{-1.0cm} \caption {(Color online) Based on the variational
method (V), in-phase and out-of-phase breathing mode dispersions
are plotted as a function of $a_{12}$ for a two-component boson
system. In each frame, first turning point in lower curves
corresponds to where the system starts to be phase separated. The
second turning point corresponds to the exchange of the in-phase
and out-of-phase modes. In the left frame of $N=10^4$, the results
given by numerical calculation [19] (N) are compared. The two
crosses at $a_{12}=0$ correspond to the breathing modes of
individual component 1 and 2 (decoupled regime).} \label{fig7}
\end{figure}

Substitution of the above dynamical wave function into
Eqs.~(\ref{eq:T1}) and (\ref{eq:T2}), one can derive the coupled
equations of motion [using (\ref{eq:T3})] for $\varepsilon_i$ and
$\varepsilon_i^\prime$, and then solve a corresponding dynamical
matrix to obtain the dispersions of relations for the breathing
modes. In Fig.~\ref{fig7}, we present the breathing mode
dispersion as a function of the inter-component scattering length
$a_{12}$. Again, all parameters are the same as before. In the
case of $N=10^4$, the results given by numerical calculation in
Ref.~\cite{pu98} is included for comparison.

Much information is revealed in Fig.~\ref{fig7}. Taking the
$N=10^4$ case as an example, when interspecies scattering length
$a_{12}$ is negative, the system is miscible. In-phase breathing
mode has lower energy and is more easily excited. When $a_{12}$ is
positive, the results are divided into three regimes. When $0\leq
a_{12}\alt 2.5$, the system is miscible and thus out-of-phase
breathing mode has lower energy and more easily gets excited. When
the system starts to phase separate but most parts of the two
components are still overlapping ($2.5\alt a_{12}\alt 4$), the
energy of the out-of-phase breathing mode remains lower than that
of in-phase mode . When the two components are well separated (not
much overlap) at $a_{12}\agt 4$ (see also Fig.~5), in-phase
breathing mode turns out to have lower energy again. The switch of
the in-phase and out-of-phase breathing modes in the above latter
two cases can be understood alternatively as follows. When phase
separation occurs but the two components are still very much
overlapped, out-of-phase mode is more easily excited because it
corresponds to the decrease of the overlap (and the energy).
However, when the two components are well separated, in-phase mode
is in turn more easily excited because in this case out-of-phase
mode simply increases the overlap (and the energy).

As seen in Fig.~\ref{fig7}(a), there are two turning points in the
lower dispersion curve. The first turning point is the signature
of the system being phase separated. While the second turning
point corresponds to the exchange of the in-phase and out-of-phase
breathing modes in the regime where the two components are well
separated. In comparison with the numerical results given by
Ref.~[19] (where only the lowest-energy mode is given), our result
of lowest-energy mode matches theirs quantitatively before phase
separation arises. After the phase separation occurs, our result
is qualitatively similar to their. (There is no second turning
point seen in Ref.~[19] for the range plotted.) It is worth noting
that the two points of $a_{12}=0$ correspond actually to the
breathing modes of two decoupled components (lower-frequency one
is for Rb, while higher-frequency one is for Na). From the
goodness of the match of our results to those of Ref.~[19], it
strongly suggests that modify Gaussian is also a good function to
describe the dynamics of a one-component system when the nonlinear
effect is important.

\section{Conclusions} \label{sec5}

In conclusion, the density profile, the ground-state energy, and
the breathing modes of a two-component Bose condensed system are
studied using a variational method. We propose a modified Gaussian
distribution function which is shown to be more effective for a
two-component system. While the modified Gaussian function
reserves the advantage of easy integration with the simple
Gaussian, it actually gives a better description even to the
equilibrium and dynamical properties of a single-component system.
Our approach can be generalized to study the vortices of a
two-component system, and possibly the system of imbalanced spin
population.


\acknowledgements We acknowledge fruitful discussions with Daw-Wei
Wang. This work is supported by National Science Council of Taiwan
under Grant No. 94-2112-M-003-011. We also acknowledge the support
from the National Center for Theoretical Sciences, Taiwan.



\end{document}